\documentclass[a4paper,12pt]{article}
\usepackage{amssymb}
\usepackage{cite}

\def\beq{\begin{equation}}
\def\eeq{\end{equation}}
\def\bea{\begin{eqnarray}}
\def\eea{\end{eqnarray}}
\def\nn{\nonumber}
\def\bra#1{\left\langle #1\right|}
\def\ket#1{\left| #1\right\rangle}
\def\braket#1#2{\langle #1 | #2 \rangle}
\def\qn#1{ \left[#1 \right]_q }
\def\hqn#1{ \left[#1 \right]_{\sqrt{q}} }
\def\T2{T^2_F}
\begin{document}
\thispagestyle{empty}

\vspace*{3cm}

\begin{center}
{\LARGE\sf
  Fuzzy Torus via $q$-Parafermion
}

\bigskip\bigskip
N. Aizawa

\bigskip
\textit{
Department of Mathematics and Information Sciences, \\
Graduate School of Science, 
Osaka Prefecture University, \\
Daisen Campus, Sakai, Osaka 590-0035, Japan}\\

\bigskip
and

\bigskip
R. Chakrabarti \\
\bigskip
\textit{
Department of Theoretical Physics, \\
University of Madras, \\
Guindy Campus, Chennai 600 025, India
}

\end{center}

\vfill
\begin{abstract}
We note that the recently introduced fuzzy torus  
can be regarded as a $q$-deformed parafermion. Based on this picture, 
classification of the Hermitian representations of the fuzzy torus is 
carried out. The result involves Fock-type representations and new 
finite dimensional representations for $q$ being a root of unity as 
well as already known finite dimensional ones. 
\end{abstract}

%
%
%
\clearpage
\setcounter{page}{1}
\section{Introduction}
\label{intor}

  Fuzzy spaces have been widely accepted as models of 
noncommutative manifolds in the context of quantum field theory  
(see for example \cite{BKV}). The fuzzy 2-sphere \cite{Mad} 
has been studied extensively. Regarding the ordinary 2-sphere 
$ S^2 $ as an one-dimensional complex projective space 
$ {\mathbb C}P^1, $ construction of the fuzzy $ S^2 $  
has been extended to higher dimensional fuzzy projective spaces \cite{BDLM}. 
Many of the fuzzy spaces obtained so far are related to the complex 
projective space $  {\mathbb C}P^n. $ Supersymmetric extensions of the 
fuzzy $ S^2 $ \cite{GR} and the fuzzy $ {\mathbb C}P^n $ \cite{IMT} 
embody such relations. Exploiting the observation that 
$ {\mathbb C}P^3 $ is an $ S^2 $ bundle over $ S^4$, the  fuzzy $ S^4 $ 
has been  constructed \cite{MOC,Abe}. Fuzzy $S^5$ has been established 
by utilizing a $ U(1) $ fibration over a fuzzy ${\mathbb C}P^2$
\cite{LRG}. Further examples are found in the investigations of fuzzy 
unitary Grassmannian spaces \cite{DJ,BBEW} and  fuzzy orthogonal 
Grassmannian spaces \cite{DOCP}. Furthermore,  fuzzy versions of various 
toric varieties have been considered in \cite{Sa} by embedding such 
varieties in $ {\mathbb C}P^n. $ 

 The sphere $ S^2 $ is a Riemann surface of genus zero that may be 
embedded in $ {\mathbb R}^3. $ Using the Cartesian coordinates 
$(x, y, z)$ in $ {\mathbb R}^3 $ it is described by a constrained 
polynomial $ C(x,y,z) = 0, $ where $ C(x,y,z) = x^2 + y^2 + z^2 -1. $ 
Similar polynomial description of compact connected Riemann surfaces of 
arbitrary genus has been introduced in \cite{ABHHS}. These authors have 
developed a general recipe for constructing fuzzy Riemann surfaces. 
Explicit realization for the torus (genus one) is completed by defining 
the fuzzy torus as a nonlinear $C$-algebra with three generators 
\cite{ABHHS}. The fuzzy torus algebra $ \T2 $ has been recast as a 
$q$-deformed Lie algebra in \cite{Naka}. This may be understood as a 
linearization of the nonlinear $ \T2 $ algebra by using the 
$q$-deformed commutators. 

  In the present work, we are looking at the fuzzy torus algebra 
from a different viewpoint. We regard $ \T2 $ as a $q$-deformed 
parafermion algebra which is different from the $q$-deformations of 
parafermion or paraboson discussed so far 
\cite{Pal,FV,CPT,Pal2,Had,BD}. Using this approach, we classify the 
Hermitian representations of $ \T2. $ In addition to recovering the 
known finite dimensional representations \cite{ABHHS,Naka}, some new 
and nontrivial representations are found. A Fock-type infinite 
dimensional representation, and a finite dimensional representation 
for $q$ being a root of unity are produced. The classification of 
the Hermitian representations will facilitate  deeper understanding of 
$ \T2 $ and model building of quantum field theory on the fuzzy torus. 
In the next section, we recall the definition of $ \T2 $ and 
explain the interpretation as $q$-deformed parafermion. Representations 
of $ \T2 $ are systematically constructed under simple assumptions in 
\S \ref{HR} so as to achieve the classification. Our concluding remarks 
are given in \S \ref{CR}.

%
%
%
\section{Fuzzy torus and deformed parafermion}
\label{PF}

   The polynomial description \cite{ABHHS} of the torus comprising of 
three real commuting variables reads
\beq
  C(x,y,z) = (x^2+y^2 - \mu)^2 + z^2 -c,  
  \label{torus_pol}
\eeq
where $ \mu $ and $ c\, (> 0)$ are real parameters. In \cite{ABHHS}, 
the parameter $c$ is set equal to unity. Following \cite{Naka}, we, 
however, keep its value arbitrary. 
The relation $ C(x,y,z) = 0 $ describes a surface in $ {\mathbb R}^3, $ 
which is a squashed sphere for the domain $- \sqrt{c} < \mu < \sqrt{c}$, 
and a torus for $\sqrt{c} < \mu. $ 

  Fuzzy analog of the surface is introduced by replacing the commuting 
variables $(x,y,z)$ with the generating elements $ (X, Y, Z)$ that 
satisfy the commutation relations \cite{ABHHS}
\bea
      [X, Y]  = i\hbar Z, \qquad
      [Y, Z] &=& i\hbar \{X, \varphi \}, \qquad
      [Z, X]  = i\hbar \{Y, \varphi \},
   \label{DefTorus} \\
      \varphi & \equiv &  X^2 + Y^2 - \mu, \label{phi}
\eea
and the constraint
\beq
  C_{F} \equiv \varphi^2 + Z^2 = c. \label{CasimirTorus}
\eeq
The parameter $ \hbar $ imparts noncommutativity. Its commutative limit 
is given by $\hbar \rightarrow 0$. The associative algebra $ \T2$,  
generated by the variables $(X, Y, Z)$ obeying the relations 
(\ref{DefTorus}, \ref{phi}), satisfy the Jacobi identity. The 
element $ C_F $ is the center of the algebra.

 Introducing the complex variable $ W = X + iY $ and its Hermitian 
adjoint $W^{\dagger}$, the defining relations (\ref{DefTorus}, 
\ref{phi}) are reexpressed as
\bea
  [W, W^{\dagger}] = 2\hbar Z, \qquad
  [Z, W] &=& \hbar \{ W, \varphi \}, \qquad
  [Z, W^{\dagger}] = -\hbar \{ W^{\dagger}, \varphi \},
  \label{WWZ_rel} \\
  \varphi & \equiv & \frac{1}{2} \{ W, W^{\dagger} \} - \mu.
\eea
Eliminating $ Z $ via the first equation in (\ref{WWZ_rel}), it is 
observed \cite{ABHHS} that the operators $W, W^{\dagger}$ obey the 
following trilinear relation and its Hermitian adjoint:
\beq
  W (W^{\dagger})^2 -2 \frac{1-\hbar^2}{1+\hbar^2} W^{\dagger} W W^{\dagger}  + (W^{\dagger})^2 W 
  = \mu \frac{4\hbar^2}{1+\hbar^2} W^{\dagger}.
  \label{Trilinear}
\eeq
The deformation parameter $q$ is introduced \cite{Naka} by the relation
\beq
  2 \frac{1-\hbar^2}{1+\hbar^2} = q + q^{-1}.        \label{q_def}
\eeq
It follows that the complex parameter $q$ is of unit magnitude: 
$ |q| = 1. $ The commutative $\hbar \rightarrow 0$ limit now corresponds 
to $q \rightarrow 1$. We now scale the conjugate variables $W$ and 
$W^{\dagger}$ as follows
\beq
  a = \left(\frac{2}{|\mu| (2 - q-q^{-1})} \right)^{1/2} W, 
  \qquad
  a^{\dagger} = \left(\frac{2}{|\mu| (2 - q-q^{-1})} \right)^{1/2} 
  W^{\dagger}.
  \label{mu_mag}
\eeq
For a fixed $\mu$ above scaling is well-defined in the 
\textit {noncommutative} 
$q \neq 1$ \textit {i.e.} $\hbar \neq 0$ regime. 
The trilinear relation (\ref{Trilinear}) now assumes two distinct 
forms depending on the sign of the parameter $\mu$. For the choice 
$\mu < 0$, it reads
\beq
 a(a^{\dagger})^2 - \qn{2} a^{\dagger} a a^{\dagger} 
 + (a^{\dagger})^2 a = -2 a^{\dagger},
 \label{Tri_a_nega}
\eeq
where the $q$-number is defined as usual
\beq
  \qn{x} = \frac{q^x - q^{-x}}{q-q^{-1}}.
\eeq
For the alternate $ \mu > 0$ case, the relation (\ref{Trilinear}) may 
be recast as 
\beq
 a(a^{\dagger})^2 - \qn{2} a^{\dagger} a a^{\dagger} 
 + (a^{\dagger})^2 a = 2 a^{\dagger}.
 \label{Tri_a_posi}
\eeq

The trilinear relation (\ref{Tri_a_nega}) may be regarded as 
$q$-deformation of the parafermion \cite{OK}. 
The relation (\ref{Tri_a_posi}) has the same form albeit 
with an opposite sign on the right hand side. One may regard this as 
a variant of the $q$-deformed parafermion. It becomes evident in the 
next section that the $q$-parafermionic picture observed in 
(\ref{Tri_a_nega}) and (\ref{Tri_a_posi}) facilitates a systematic 
study of the representations. The transformation (\ref{mu_mag}) is 
singular for a fixed $\mu$ in the limit $q \rightarrow 1$. In this 
scenario the classical $q = 1$ parafermions do not arise. There is, 
however, an interesting possibility 
\beq
|\mu| \sim (2 - q -q^{-1})^{-1},
\label{class_lim}
\eeq
which allows taking the $q \rightarrow 1$ limit in (\ref{mu_mag}). 
This limit is pertinent in understanding some of the 
representations discussed below.

%
%
%
\setcounter{equation}{0}
\section{Hermitian representations}
\label{HR}

   In this section, we obtain all Hermitian representations of 
the algebra defined by (\ref{DefTorus}). 
Our strategy is as follows: we start with the smallest number of 
assumptions and construct the most general representation. 
Then, further assumptions are imposed to obtain possible 
subclasses of the representations. As  observed in \cite{ABHHS}, 
the operators $ aa^{\dagger} $ and $ a^{\dagger} a $ commute, and,
therefore, they are simultaneously diagonalizable. We thus assume 
the existence of a normalized state $ \ket{0} $ such that
\beq
   aa^{\dagger} \ket{0} = p \ket{0}, \qquad
   a^{\dagger} a \ket{0} = r \ket{0}, \qquad
   \braket{0}{0} = 1. \label{vacuum}
\eeq
Other states may be obtained by repeated actions of $ a $ or 
$ a^{\dagger} $ on $ \ket{0}. $ The states are required to be 
normalizable: 
\beq
   \ket{n} = {\cal N}_n (a^{\dagger})^n \ket{0}, \qquad
   \ket{-n} = {\cal N}_{-n} a^n \ket{0} 
   \quad \forall \;\; n  > 0, \label{states}
\eeq
where $ {\cal N}_{\pm n} $ are the normalization constants. 
Normalization of the $ \ket{\pm 1} $ states puts constraints on the 
values of $p$ and $r$: 
\[
 \| \ket{1} \|^2 = |{\cal N}_1|^2 \bra{0} aa^{\dagger} \ket{0} 
 = |{\cal N}_1|^2 p,
 \qquad
 \| \ket{-1} \|^2 = |{\cal N}_{-1}|^2 \bra{0} a^{\dagger} a \ket{0} 
 = |{\cal N}_{-1}|^2 r.   
\]
It follows that $ p, r > 0 $, and we choose 
$ {\cal N}_1 = p^{-1/2}, \ {\cal N}_{-1} = r^{-1/2}.$ For the states of 
$ {\ket {\pm n} \forall n \geq 2}, $ we use the relations 
(\ref{Tri_a_nega}) or (\ref{Tri_a_posi}) appropriately for evaluating 
the norm. For the regime $ \mu < 0, $ the following relations are 
verified inductively:
\bea
  & & a \ket{n} = 
     \left( \qn{n} p - \qn{n-1} r - 2 \qn{\frac{n-1}{2}} \hqn{n} \right)^{1/2} \ket{n-1}
     \equiv A_{n} \ket{n-1}, \label{An} \\ 
  & & a^{\dagger} \ket{n} = A_{n+1} \ket{n+1}, \label{An1}
\eea
and
\beq
  a \ket{-n} = A_{-n} \ket{-n-1}, \qquad
  a^{\dagger} \ket{-n} = A_{-n+1} \ket{-n+1}.
  \label{Amn}
\eeq
The diagonal operators read
\beq
  a^{\dagger} a \ket{\pm n} = A_{\pm n}^2 \ket{\pm n}, \qquad
  a a^{\dagger} \ket{\pm n} = A_{\pm n+1}^2 \ket{\pm n}.  \label{Diagonal}
\eeq
For the choice $ \mu > 0 $ the construction of states proceeds as 
before:
\bea
  & & a \ket{n} = 
     \left( \qn{n} p - \qn{n-1} r + 2 \qn{\frac{n-1}{2}} \hqn{n} \right)^{1/2} \ket{n-1}
     \equiv {\mathsf A}_{n} \ket{n-1}, \label{An_posi} \\
  & & a^{\dagger} \ket{n} = {\mathsf A}_{n+1} \ket{n+1}, \label{An1_posi}
\eea
while the relations (\ref{Amn}) and (\ref{Diagonal}) hold with the 
replacement of the normalization constant $ A_n $ by $ {\mathsf A}_n.$
The Casimir operator is quartic in the variables $ W, \; W^{\dagger}.$  
Its eigenvalues may be computed via the results obtained above: 
\beq
  \frac{4}{\mu^2 \, (2-\qn{2})}\, C_F \ket{\pm n} = 
  \left\{
    \begin{array}{lcl}
      {\displaystyle 
      \left(
        p^2+r^2 - \qn{2} pr + 2(p+r) + \frac{4}{2-\qn{2}}
      \right) \ket{\pm n},
      }
       &  & \mu < 0, \\[20pt]
      {\displaystyle 
      \left(
        p^2+r^2 - \qn{2} pr - 2(p+r) + \frac{4}{2-\qn{2}}
      \right) \ket{\pm n},
      }
       &   & \mu > 0.
    \end{array}
  \right.
  \label{CFeigen}
\eeq
We have \textit {formally} obtained a double-sided infinite dimensional 
representation with two parameters: $ p, r.$ Our representation has 
close kinship with the symplecton realization \cite{BL71} of the boson 
calculus. Reflecting the symmetry of (\ref{Tri_a_nega}) 
(or (\ref{Tri_a_posi})) and its adjoint under the exchange of $a$ and 
$a^{\dagger},$ the representation has the symmetry structure 
$A_{n+1} (p, r) = A_{-n} (r, p)$. The usefulness of the above 
representations becomes evident below where we impose restrictions for 
obtaining the subclasses of the above infinite dimensional representation. 
These restrictions terminate the infinite series of state vectors. 

   Assuming that the state $ \ket{0} $ is annihilated by the operator
$ a, $ we have, $ r = 0. $ All states $ \ket{-n} $ labeled by negative 
integers are eliminated while the semi-infinite series of states 
$\{\ket{n} | n= 0, 1, 2, \dots \infty\}$ are retained. This is a Fock-type 
representation constructed on the vacuum $ \ket{0}. $ The number of 
parameters is reduced to one. 

A key requirement for the existence for the classes of representations 
discussed below is that their \textit {Hermiticity needs to be 
maintained}. This, in turn, requires the constants 
$A_{\pm n}^{2}, {\sf A}_{\pm n}^{2} \forall n > 0$ to be 
\textit {nonnegative}. Keeping this in mind, we study all the 
possibilities in order.

\medskip\noindent
{\bf (1) Truncated Fock-type representation}

To obtain finite dimensional Fock-type representations, we further 
assume that there exist a positive integer $N$ such that 
\beq
    A_N = 0\;\;\; \hbox{for} \;\;\;\mu < 0, \qquad   
    {\mathsf A}_N = 0 \;\;\;\hbox{for} \;\;\;\mu > 0. \label{fin_cond}
\eeq
The representation space is 
spanned by $N$ independent states: $\ket{0}, \ket{1}, \dots, \ket{N-1}$
with the highest state satisfying $ a^{\dagger} \ket{N-1} = 0. $ 
For instance, the $N$ dimensional representation for $\mu < 0$ reads
\beq
  a^{\dagger} = 
  \left(
    \begin{array}{ccccc}
      0 & A_{N-1} & 0 &  \cdots & 0  \\
        & 0   & A_{N-2} &  \cdots & 0 \\
        &     & \ddots & \ddots  & \vdots \\
        &     &        &   0     & A_1 \\
        &     &        &         & 0
    \end{array}
  \right).
  \label{string}
\eeq
The null condition (\ref{fin_cond}) requires
\beq
   \qn{N} p \mp 2\qn{\frac{N-1}{2}} \hqn{N} = 0,   \label{AN0}
\eeq
where the upper (lower) sign refers to $\mu < 0\; (\mu > 0)$. The 
relation (\ref{AN0}) holds for two cases:
 
({\sf i}) A generic value of 
\beq 
q  = \exp (i \theta) 
\label{q_gen}
\eeq
leads to a $N$-dependent order parameter $ p$: 
  \beq
     p = \pm 2 \qn{\frac{N-1}{2}} \frac{ \hqn{N} }{ \qn{N} } 
     = \pm \left(\frac{\tan (N \theta /2)}{\tan (\theta/2)} - 1
     \right). \label{pN}
  \eeq
This case corresponds to the `string solutions' of \cite{ABHHS,Naka}. 
The angle $\theta$ is restricted by the requirement of nonnegativity 
of the elements $A_{n}^{2}\: (\mathsf{A}_{n}^{2})\: 
\forall n \in (1, \dots, N - 1)$:
\beq
     A_{n}^{2} = - \mathsf {A}_{n}^{2}
     = \frac{\sin^{2}(n \theta /2)}{\sin^{2} (\theta/2)}
     \left(\frac{\tan (N \theta /2)}{\tan (n \theta/2)} - 1\right).
     \label{A_n_square}
\eeq
For the regime $\mu < 0$ the positivity $A_{n}^{2} > 0$ holds in the 
domain $- \pi /N < \theta < \pi/N$, whereas for the choice $\mu > 0$ 
the positivity $\mathsf{A}_{n}^{2} > 0$ is satisfied, for instance,  
in the restricted region $\pi /N < \theta < \pi/(N - 1), \: 
- \pi /(N - 1) < \theta < - \pi/N$. This class of representation 
matrices are symmetric with respect to the minor diagonal: 
$ A_{N-k} = A_k,\, {\mathsf A}_{N - k} = {\mathsf A}_{k}.$ 
 
({\sf ii}) The root of unity values of $q = \exp(i 2\pi/N) $ restricts
$ p \,(> 0) $\, over a range. The solutions corresponding to this class is 
novel. We list some low dimensional Hermitian representations for 
$\mu > 0$ as examples: 
\[
  a^{\dagger} = 
  \left(
    \begin{array}{ccc}
      0 & \sqrt{2 - p} & 0 \\
        &    0     & \sqrt{p} \\
        &          & 0
    \end{array}
  \right),
  \quad
  a^{\dagger} = 
  \left(
    \begin{array}{cccc}
      0 & \sqrt{2 - p} & 0 & 0 \\
        & 0  & \sqrt{2} & 0 \\
        &    &  0 & \sqrt{p} \\
        &    &    &  0
    \end{array}
  \right),
\]
\beq
  a^{\dagger} = 
  \left(
    \begin{array}{ccccc}
      0 & \sqrt{2 - p} & 0 & 0 & 0 \\
        &    0        &  \sqrt{2 (1 + (2 - p)\, \sin \frac{\pi}{10})} 
        & 0 & 0 \\
        &             &    0  &  \sqrt{2 (1 + p\, \sin \frac{\pi}{10})} 
        & 0 \\
        &             &       &   0  & \sqrt{p} \\
        &             &       &      &    0
    \end{array}
  \right).
\label{unity_rep}
\eeq
The elements $\mathsf {A}^{2}_{n} \, \forall n \in (1, \dots, N - 1)$ 
for the representations (\ref{unity_rep}) read 
\beq 
\mathsf{A}^{2}_{n} = \frac{\sin^{2}(n\,\pi/N)}{\sin^{2}(\pi/N)}
\;(1 + (p - 1)\,\mathsf {f}_{n}), \qquad
\mathsf{f}_{n} = \frac{\tan (\pi/N)}{\tan (n\,\pi/N)}.
\label{rt_unit_herm}
\eeq
The sequence $\{\mathsf{f}_{n}| n =1,\dots, N - 1\}$ is bounded as 
$- 1 \leq \mathsf{f}_{n} \leq 1$.  
The Hermiticity requirement now restricts the order parameter as 
$0 < p < 2$. The value of $p$ being positive definite, 
the Hermitian  representations of this class do not exist for $\mu < 0$. 

\medskip\noindent
{\bf (2) Cyclic representations}

The cyclic finite dimensional representations may be ensured by 
identifying the states $ \ket{N} $ and $ \ket{0}. $ In particular, this 
makes the eigenvalues of the operators $aa^{\dagger}$ and $a^{\dagger}a$  
on the states $ \ket{N} $ and $ \ket{0} $ same. For the $\mu > 0$ case, 
we obtain
\beq
    \mathsf{A}_{N+1}^2 = p, \qquad \mathsf{A}_N^2 = r. 
    \label{cyclic_cond}
\eeq
Eliminating $r$ from the relations in (\ref{cyclic_cond}), we obtain
\[
    \hqn{N}^2\, p =  \frac{2}{2 - \qn{2}} \hqn{N}^2.
\]
For the choice $\hqn{N} \neq 0,$ it follows $p = 2 (2 - \qn{2})^{-1}.$ 
It then turns out that $r = p$, and $\mathsf{A}_n = \sqrt{p}\;\;
\forall \, n$. This leads to $ Z \ket{n} = 0 $ \cite{ABHHS}. This case is 
uninteresting. Alternate choice $ \hqn{N} = 0, \ i.e. \ q^N = 1$ 
yields  a class of two-parametric $(p, r) \: N$-dimensional 
representations:
\beq
  a^{\dagger} = 
  \left(
    \begin{array}{ccc}
      0 & \sqrt{2 - p - r} & 0 \\
        &    0     & \sqrt{p} \\
      \sqrt{r}  &          & 0
    \end{array}
  \right),
  \quad
  a^{\dagger} = 
  \left(
    \begin{array}{cccc}
      0 & \sqrt{2 - p} & 0 & 0 \\
        & 0  & \sqrt{2 - r} & 0 \\
        &    &  0 & \sqrt{p} \\
      \sqrt{r}  &    &    &  0
    \end{array}
  \right),
\label{cyclic_rep}
\eeq
\[
  a^{\dagger} = 
  \left(
    \begin{array}{ccccc}
      0 & \sqrt{2 ( 1 + r \,\sin \frac{\pi}{10}) - p} & 0 & 0 & 0 \\
        &    0       &  \sqrt{2 (1 + (2 - p - r)\, \sin \frac{\pi}{10})} 
        & 0 & 0 \\
        &            &   0 &  \sqrt{2 (1 + p\, \sin \frac{\pi}{10}) - r} 
        & 0 \\
        &            &       &   0  & \sqrt{p} \\
       \sqrt{r} &            &       &      &    0
    \end{array}
  \right).
\]
These representations corresponds to the `loop solutions' in 
\cite{ABHHS,Naka}, and possess the symmetry 
\beq
{\mathsf A}_{n} (p, r) = {\mathsf A}_{N + 1 - n} (r, p). 
\label{pr_sym}
\eeq
To examine their nonnegativity the elements 
$\{{\mathsf A}_{n}^{2}| n \in (1, \dots, N)\}$ may be recast by 
isolating their symmetric part ${\cal S}_{n}(p + r)$ as follows 
\beq
{\mathsf A}_{n}^{2} = {\cal P}_{n} \, p + {\cal S}_{n}(p + r),\;\;
{\cal P}_{n} = [n]_{q} + [n - 1]_{q}, \;\;
{\cal S}_{n}({\cal X}) = - [n - 1]_{q}\, {\cal X} + 2\,
\Big[\frac{n - 1}{2}\Big]_{q}\,[n]_{\sqrt{q}}.
\label{APS}
\eeq
Due to the symmetry (\ref{pr_sym}) we only need to check the 
nonnegativity of the elements ${\mathsf A}_{n}^{2}$ for 
$n = 1, \dots, N/2\;\; ((N + 1)/2)$ for an even (odd) $N$. 
In this domain we have ${\cal P}_{n} \geq 0$, and
\beq
{\cal S}_{n}(p + r) = \frac{\sin^{2}((n - 1) \pi/ N)}   
{\sin^{2}(\pi/ N)}\, (1 + (1 - p - r)\, {\mathsf f}_{n - 1}),
\label{S_val}
\eeq
where the sequence $\{{\mathsf f}_{n}\}$ has been defined in 
(\ref{rt_unit_herm}). In the present context the entries of the sequence 
$\{{\mathsf f}_{n - 1}\}$, where $n = 2, \dots, N/2\; ((N+ 1)/2)$ for 
an even (odd) $N$, is bounded as 
$0 < {\mathsf f}_{n - 1} \leq 1$. This yields the parametric values 
$0 < p, 0 < r, p + r < 2$ for the required nonnegativity. 
For the odd-$N$ case the symmetric element reads
\beq
{\mathsf A}_{\frac{N + 1}{2}}^{2} = \frac{1}{2\, \cos (\pi/N)}\,
\Big(\frac{1}{2\, \sin^{2} (\pi/ 2 N)} - p - r\Big ).
\label{symval_A}
\eeq
It may be noticed that the solutions obtained in (\ref{unity_rep}) may 
be obtained as the $r \rightarrow 0$ limiting case of the present cyclic 
representations. The corresponding matrix structures are, however, of 
different ranks, and in that sense inequivalent. Moreover the cyclic 
property of the solutions (\ref{cyclic_rep})\, is \textit {lost} when 
$r = 0$ is substituted in them. Hermitian cyclic representations do not 
exist for the $\mu < 0$ case as the elements $\{A_{n}^{2} | n = 1,\dots,N\}$ 
are not nonnegative. For instance, we have
\beq
A_{\frac{N}{2}}^{2}\Big|_{\hbox{even}\;\; N} 
= - r - \frac{1}{\sin^{2} (\pi/ N)}, \qquad 
A_{\frac{N + 1}{2}}^{2}\Big|_{\hbox{odd}\;\; N} 
= -\frac{1}{2 \cos (\pi/ N)}\;
\Big(\frac{1}{2\,\sin^{2}(\pi/2 N)} + p + r\Big) . 
\label{muneg_no_cyclic}
\eeq

\medskip\noindent
{\bf (3) Infinite dimensional representations}

Turning towards the infinite dimensional representations we discuss
the special cases where we notice the existence of such Hermitian 
representations. The identity 
\[
2\,\Big[\frac{n - 1} {2}\Big]_{q}\, [n]_{\sqrt{q}} = [n]^{2}_{\sqrt{q}}
- [n]_{q}
\]
allows us to rewrite the elements ${\mathsf A}_{n}^{2}$ for the Fock 
states $(r = 0)$ in the $\mu > 0$ case as
\beq
{\mathsf A}_{n}^{2} = [n]_{\sqrt{q}}^{2} + (p - 1)\, [n]_{q}.
\label{Fock_A}
\eeq
For the choice $p = 1$, therefore, the Hermiticity of the infinite set of 
basis states with $n = 1, 2, \dots \infty $ is guaranteed. To establish a 
range of values of the order parameter $p$ around $p = 1$ that preserves 
the positivity ${\mathsf A}_{n}^{2} > 0$,  we use 
(\ref{q_gen}) to 
obtain
\beq 
{\mathsf A}_{n}^{2} = \frac{\sin^{2}(n \theta/ 2)}{\sin^{2}(\theta/2)}\,
(1 + (p - 1)\, \mathfrak{f}_{n}),\qquad
\mathfrak{f}_{n} = \frac{\tan (\theta/2)}{\tan\, (n \theta /2)}.
\label{Fock_A_theata}
\eeq 
The entries of the sequence 
$\{\mathfrak{f}_{n} | n = 1, 2, \dots \infty\}$ for any positive 
\textit{integral} value of $n$ remain \textit{finite} for the choice
$\theta = 2 \pi/ {\cal N}$, where ${\cal N}$ is an \textit{irrational}
number. Employing the finite maximum and minimum entries of the said 
sequence a range of values of $p$ may now be easily established for 
which Fock-state representations are Hermitian and well-defined.

Another possible scenario is to turn to the quasi-classical states
mentioned in the context of (\ref{class_lim}). Using the classical limit 
$q \rightarrow 1$ in the context of the $(p, r)$ two-parametric states
for the regime $\mu > 0$ we obtain
\beq
{\mathsf A}_{n}^{2} = n^{2} + (p - r - 1) \, n + r 
\quad \forall \,n = 1, 2, \dots \infty.
\label{class_A}
\eeq
For the choice $ p \geq r + 1$, these two-parametric states maintain 
${\mathsf A}_{n}^{2} > 0$ ensuring the Hermiticity of the representations. 
These infinite dimensional quasi-classical states reflect, in some sense, 
noncommutative spaces bordering on the commutative regime. Hermitian 
infinite dimensional representations for the $\mu < 0$ case do not exist.
 
\bigskip
  We have studied so far the case of $ \mu \neq 0. $ 
To complete the study of Hermitian representations, 
we next investigate the case of  $ \mu = 0. $ 
Although the algebra of this case does not correspond to 
a deformed parafermion, one can repeat the analysis used earlier. 
We use the trilinear relation
\beq
   W (W^{\dagger})^2 - \qn{2} W^{\dagger} W W^{\dagger} + (W^{\dagger})^2 W = 0, 
   \label{mu0cubic}
\eeq 
and assume a state $ \ket{0} $ subject to the conditions
\beq
  WW^{\dagger} \ket{0} = p \ket{0}, \qquad 
  W^{\dagger} W \ket{0} = r \ket{0}, \qquad
  \braket{0}{0} =1.  \label{mu0vac}
\eeq
Using the same arguments applied earlier the positivity   
$ p, r > 0 $  follows, and the states are given by 
\bea
  & & W \ket{n} = \left( \qn{n} p - \qn{n-1} r \right)^{1/2} \ket{n-1} \equiv 
      w_n \ket{n-1},
      \label{Wn} \\
  & & W^{\dagger} \ket{n} = w_{n+1} \ket{n+1}.  \label{Wdagn}
\eea
This leads to
\beq
   W^{\dagger} W \ket{n} = w_n^2 \ket{n}, \qquad
   W W^{\dagger} \ket{n} = w_{n+1}^2 \ket{n}.
   \label{mu0diag}
\eeq
The eigenvalue of the Casimir operator reads
\beq
  C_F \ket{n} = \frac{p^2+r^2 - \qn{2}pr}{2 -\qn{2}} \ket{n}.
  \label{mu0Cas}
\eeq
Relations parallel to (\ref{Wn}) - (\ref{mu0Cas}) also hold for the 
state $\ket{-n}. $ 

  Fock-type representation is obtained by requiring $ W \ket{0} = 0 $ 
\textit {i.e.} $ r = 0. $ Implementing $ w_N = 0 $ we may truncate the 
Fock-type representation and obtain a finite $N$ dimensional one. 
There is a slight difference from the case of $ \mu \neq 0. $ 
The equation
\[
  w_N  = \sqrt{ \qn{N} p } = 0, \quad p > 0
\]
may be solved only for the choice $ \qn{N} = 0$. Namely, it is possible 
to obtain finite dimensional Hermitian representations from the Fock-type 
one for $ q^{2N} = 1$ \textit{i.e.} $q = \exp (i \pi/ N)$. The 
expression for $w_{n}^{2}$ is found to be nonnegative: 
\beq 
w_{n}^{2}= \frac{\sin (n \pi/ N)}{\sin (\pi/ N)}\;p > 0\;\;\;\forall n 
\in (1, \dots, N - 1).
\label{w_Fock}
\eeq
Such a truncation for a generic $q$ does not exist. 

  Now we investigate the possible realizations of the cyclic 
representations for the $\mu = 0$ case. Conditions for cyclic 
representation are given by
\beq
   w_N^2 =r, \qquad w_{N+1}^2 = p. \label{mu0cyc}
\eeq 
Above equations may be recast as 
\beq
   \qn{N} p = (\qn{N-1} + 1)r, \qquad
   (\qn{N+1} - 1) p = \qn{N} r. 
   \label{mu0rp}
\eeq
Eliminating $r$ from (\ref{mu0rp}) we obtain a relation for $p$:
\[
  q^{-N}(1-q^N)^2 p = 0. 
\]
Therefore the cyclic representations may exist only if $q^N = 1$.
By computation, however, it may be seen that negative values of 
$w_{n}^{2}$ exist:
\beq
w_{\frac{N}{2}}^{2}\Big|_{\hbox{even}\;\; N} = - r, \qquad 
w_{\frac{N + 1}{2}}^{2}\Big|_{\hbox{odd}\;\; N} 
= -\frac{p + r}{2 \cos (\pi/ N)}. 
\label{no_cyclic}
\eeq
This eliminates the possibility of having Hermitian cyclic 
representations for the $\mu = 0$ case.

Lastly, paralleling the $\mu > 0$ case the infinite dimensional 
representations for the choice $\mu = 0$ exist for the said 
semi-classical states in the $q \rightarrow 1$ limit as the elements
\beq
w_{n}^{2} \Big|_{q \rightarrow 1} = n (p - r) + r 
\quad \forall n = 1, 2, \dots \infty
\label{mu_null_inf}
\eeq
are nonnegative for $p \geq r$. 
%
%
%
\section{Concluding remarks}
\label{CR}
\setcounter{equation}{0}

  We introduced a viewpoint that regards fuzzy torus algebra as 
a $q$-deformed parafermion. Based on this picture, we investigated 
Hermitian representations of the fuzzy torus under the assumptions 
(\ref{vacuum}) and normalizability of the representation basis. 
All known representations in \cite{ABHHS,Naka} were recovered and 
some new representations (both finite and infinite dimensional) were 
discovered. To summarize, the finite dimensional Hermitian 
representations of the algebra (\ref{Tri_a_nega}) and (\ref{Tri_a_posi})
may be classified as follows. ({\sf i}) For a generic value of $q$, the
representations exist with $N$-dependent order parameter $p$ for both 
the $\mu < 0$ and $\mu > 0$ regime. ({\sf ii}) For the root of unity 
values of $q = \exp (i 2 \pi/ N)$ the Hermitian Fock-type 
representations given in (\ref{unity_rep}) exist only for the choice 
$\mu > 0$. ({\sf iii}) Two parametric cyclic representations given in 
(\ref{cyclic_rep}) exist for root of unity 
values of $q = \exp (i 2 \pi/N)$ for the $\mu > 0$ case. In the 
$r \rightarrow 0$ limit these representations reduce to 
the Fock-type solutions (\ref{unity_rep}). Nevertheless, the matrix 
structures of the two classes of solutions are of different ranks, and
the cyclic property of the solutions (\ref{cyclic_rep}) is
lost on substituting $r = 0$ in them. As these two categories of 
solutions reflect the allowed noncommutative structure of spaces, their 
geometric properties are likely to be quite distinct. 

Turning towards the existence of an infinite dimensional Hermitian 
Fock-type representation discussed here, we note that these solutions 
allow us to introduce coherent states of the fuzzy torus as eigenstates 
of the operator $ a. $ The coherent state will lead to developing a 
star-product on the fuzzy torus by the method of \cite{Vo}. This will be 
discussed in a separate publication. 

  We have seen that the fuzzy torus algebra admits interpretations 
as a $q$-deformed Lie algebra \cite{Naka}, or a $q$-deformed parafermion. 
These interpretations may not be the only possibilities. 
One can derive the following set of relations from (\ref{DefTorus}):
\bea
  & & [X, Y] = i\hbar Z, \qquad\qquad\quad [Z, \varphi] = 0, \nn \\
  & & [Z, X] = i\hbar \{ Y, \varphi \}, \qquad \quad 
      \{ Z, X \} = i\hbar^{-1} [Y, \varphi], \label{S_alg} \\
  & & [Z, Y] = -i\hbar \{ X, \varphi \}, \qquad\,
      \{ Z, Y \} = -i\hbar^{-1} [X, \varphi]. \nn
\eea
This set looks like a variant of the Sklyanin algebra \cite{Sk}. 
The implications of this viewpoint is not clear at present. 
It is also important to develop a supersymmetric extension of the 
fuzzy torus, and fuzzy surfaces of higher genus. Their relations 
with generalizations of parafermions are of interest.
%
\section*{Acknowledgements}

 The work of N.A. is partially supported by the grants-in-aid from JSPS, 
Japan (Contract No. 18540380). Other author (R.C.) is partially 
supported by DAE (BRNS), Government of India. 

%
%
%
%

\end{document}